\documentstyle [twoside,12pt]{article}
\newcommand{\nc}{\newcommand}
\nc{\la}{\lambda} \nc{\La}{\Lambda}  \nc{\al}{\alpha}
\nc{\te}{\theta}  \nc{\be}{\beta}	\nc{\ka}{\kappa}
\nc{\ga}{\gamma}  \nc{\Ga}{\Gamma}
\nc{\de}{\delta}  \nc{\De}{\Delta}
\nc{\si}{\sigma}  \nc{\Si}{\Sigma}
\nc{\om}{\omega}  \nc{\Om}{\Omega}
\nc{\nf}{\infty}   \nc{\nl}{\newline}
\nc{\ra}{\longrightarrow}
\nc{\beq}{\begin{equation}}
\nc{\eeq}{\end{equation}}
\nc{\beqa}{\begin{eqnarray}}  \nc{\dst}{\displaystyle}
\nc{\eeqa}{\end{eqnarray}} \nc{\nnb}{\nonumber}
\title{\bf Extended QED with CPT violation  : \\ clarifying
some controversies.}\author{Guy Bonneau\thanks {\noindent Laboratoire de
Physique Th\'eorique et des Hautes Energies,
 Unit\'e associ\'ee au CNRS UMR 7589, Universit\'e Paris 7,
 2 Place Jussieu, 75251 Paris Cedex 05. Email: bonneau@lpthe.jussieu.fr}}
\begin{document}
\date{}
\maketitle
\begin{abstract}
\noindent We rediscuss the controversy on a possible Chern-Simons like
term generated through radiative corrections in QED with a CPT  violating
term. We analyse some consequences of the division of the
Lagrangian density between ``free part" and ``interaction part". We also
emphasize the fact that any absence of an {\sl a priori} divergence should be
explained by some symmetry or some non-renormalisation theorem and show that
the so-called ``unambiguous result" based upon ``maximal $SO(3)$ residual
symmetry " does not offer a solution.
\end{abstract}

\vspace{5cm}



\vfill \hfill  October 2006
\newpage

\section{Introduction}
In the last decade, the interesting issue of a possible spontaneous breaking
of Lorentz invariance at low energy has been considered : this issue also led
to CPT breaking
\cite{{CFJ90},{ColKos},{ColGlas}}. In particular, the general
Lorentz-violating extension of the minimal
$SU(3)\times SU(2)\times U(1)$ standard model has been discussed : as many
breaking terms are allowed, people look for possible constraints coming from
experimental results as well as from renormalisability requirements, anomaly
cancellation, microcausality and stability \cite{{KosLeh},{Adam1},{Adam}}.

In that respect, there arose a controversy on a possible Chern-Simons like
term generated through radiative corrections (first order in the Lorentz
breakings) and on a possible mass term for the photon (second order in the
Lorentz breaking) [2-19]. In this note, we intend to clarify the origin of the
discrepancies. Among previous works on that subject, we particularly quote
\cite{{Victoria,Bonneau01,Soldati1}}. This phenomenum was
extensively  studied in QED, an abelian gauge theory, as a part of the
standard model.

 The Lagrangian density is~:
\beqa\label{QED1} {\cal L} & = & {\cal L}_0 + {\cal L}_{int} + {\cal L}_1 +
{\cal L}_2 \\ a)\ {\cal L}_0 & = & \bar{\psi}(i\not{\partial} - m)\psi -
\frac{1}{4}F_{\mu\nu}^2 -\frac{1}{2\al}(\partial A)^2 +\frac{1}{2}\la^2
A_{\mu}^2 \nnb\\
\quad & \quad & {\rm where}\ \al\ {\rm is\ the\ gauge\ parameter\ and}\
\la\ {\rm an\ infra-red\ regulator\ photon\ mass\,,} \nnb \\ b)\ {\cal
L}_{int} & = & - e\bar{\psi}\not{A})\psi \quad\quad {\rm where}\ e\ {\rm is\
the\ electron\ charge,} \nnb \\ c)\  {\cal L}_1(x) & = & -
b^{\mu}\bar{\psi}(x)\ga_{\mu}\ga^5\psi(x)\,,\quad\quad {\rm where}\ b^{\mu}\
{\rm is\ a\ fixed\ vector,}\nnb\\  d)\ {\cal L}_2(x) & = &
\frac{1}{2}c^{\mu}\epsilon_{\mu\nu\rho\si}
F^{\nu\rho}(x)A^{\si}(x)\,,\quad\quad {\rm where}\  c^{\mu}\ {\rm is\ a\
fixed\ vector.} \nnb \eeqa Other breakings could be considered (see a
discussion in the first paper of \cite{ColKos}), but we simplify and require
charge conjugation invariance, which selects
${\cal L}_1(x)$ and ${\cal L}_2(x)\,.$ Note for further reference that
experiments on the absence of birefringence of light in vacuum put very
restrictive limits on the value of $c^{\mu}\,,$ typically for a timelike
$c^{\mu}\,,\ c^{\small 0}/m \le 10^{-38 }$ \cite{ColKos}. Then, it is really
interesting to analyse the conjecture that, even if it vanishes at the tree
level, a non zero
$c^{\mu}$ might be generated through loop-corrections in presence of an ${\cal
L}_1$ term.  
Note also that the Lagrangian density
${\cal L}_0\ + {\cal L}_2$ would not lead to a coherent theory as an
(infinite) counterterm ${\cal L}_1$ appears at the one-loop order
\cite{Bonneau01}.

In \cite{Bonneau01}, the one-loop vacuum polarization tensor was computed
within the consistent dimensional scheme \cite{{t'Hooft},{DR1}} and the
complete, all-order theory was analysed in a perturbative spirit we shall
comment later on. Some other authors use non standard regularisations for
their one-loop calculations, whose consistent use in higher loop computations
is rather unclear : let us recall that a renormalisation scheme
(regularisation + subtraction algorithm + normalisation conditions) requires
delicate proofs to be consistent to all orders (recall the technicalities of
the BPHZ forest formula
\cite{BPHZ} and its specification in dimensional renormalisation in
\cite{DR1}).

However, in this note, to clarify the origin of some discrepancies, it will be
sufficient to consider the one-loop photon vacuum polarization tensor
$\Ga_{\mu\nu}(p,-p)\,.$ 
\begin{itemize}
\item As is well known, to zero'th order in the Lorentz breaking parameters,
power counting enforces a quadratic divergence, but gauge invariance
$\left[\dst p^{\mu}\Ga_{\mu\nu}(p,-p) = 0 \right] $
lowers the divergence to a logarithmic one
(the usual charge renormalisation). 

\item Then, to first order in the dimension-one Lorentz breaking parameters,
the polarization tensor should diverge linearly, but parity conservation and
gauge invariance again allows only a logarithmic one. However, {\bf all
computations agree to give a finite contribution} although everybody who
learns renormalisation theory knows that the worst always happens - except if
some extra symmetry forbids it (recall the chiral anomaly which is a finite
quantity, thanks to gauge invariance ..). It is really surprising that in the
thirty or so papers devoted to that subject, one could not find one line of
argument to explain this ``experimental " one-loop\footnote{\ If this
finiteness was an ``accidental" one (some authors remark a ``miraculous"
cancellation between two divergent quantities), it would have no reason to
hold at higher-loop order\,!} finiteness, except in the Coleman-Glashow
analysis
\cite{ColGlas} where the finiteness results from the gauge Ward identity on
{\sl the unintegrated} axial 3-point function :
\beq\label{WIl} p_{\nu}<[\bar{\psi}\gamma^{\mu}\ga^5 \psi](-p-q) A^{\nu}(p)
A^{\rho}(q)> \ =\ q_{\rho}<[\bar{\psi}\gamma^{\mu}\ga^5
\psi](-p-q) A^{\nu}(p) A^{\rho}(q)>\ =\ 0\,,
\eeq and in the review by  P\'erez-Victoria
\cite{Victoria} where the finiteness of the Chern-Simons like term is
related to that of the standard triangle graph in ordinary QED~: however,
let us recall that, here also, the finiteness results from the gauge Ward
identity on the unintegrated axial-vector-vector 3-point function
(equ.(\ref{WIl})).

\item Finally, to second order in the dimension-one Lorentz breaking parameters,
the polarization tensor should diverge logarithmicaly, but standard gauge
invariance argument forbids its appearance.

\end{itemize} In our all-order analysis, the tool of local gauge invariance
was used to prove a non-renormalisation theorem for the Chern-Simons like
term
\cite{Bonneau01}. More
recently, a series of papers \cite{Soldati1,Soldati2}, using what is called 
`` maximal
$SO(3)$ residual invariance after $SO(1,3)$ Lorentz breaking", claimed that
they offer for the first time unambiguous results for the ( {\sl still
unexplained}) finite contributions, with a non-zero induced one-loop
Chern-Simons contribution and a radiatively generated photon mass.

In this note we shall prove that the discrepancies among published results do
not come from such kind of enforcement of a symmetry, but rather from two
reasons :
\begin{itemize}
\item the main reason is the choice of the free theory : is it given by the
bilinear part $ {\cal L}_0$, such as considered by
\cite{{ColGlas},{Bonneau01},{Costa}} (we shall speak of {\sl a perturbative
approach}) or from the bilinear part of the complete lagrangian density, i.e.
$ {\cal L}_0 + {\cal L}_1 + {\cal L}_2\,,$ such as considered by
\cite{{ColKos},{Victoria99},{Soldati1}} (we shall speak of {\sl a
non-perturbative approach}) ?
\item the second one, among analyses choosing the non-perturbative approach,
lies in the choice of regularisation and renormalisation scheme. First, these
should not destroy the usual QED results (in particular the gauge invariance of
the pure QED vacuum polarisation tensor)~; second, if gauge Ward identity
holds true in the extended theory, one should prefer a regularisation that
preserves gauge invariance such as Pauli-Vilars \cite{Costa} or dimensional
regularisation
\cite{Bonneau01} or add finite quantum corrections to the Lagrangian density
to restaure the Ward identities~: in the absence of some symmetry, the
finiteness of the corrections remains unexplained and, moreover, there is no
tool to fix the ambiguous regularisation dependent values (as Jackiw said,
{\sl``When radiative corrections are finite but undetermined''}
\cite{Jackiw}).
\end{itemize}

So, in Sections 2 and 3 we successively discuss these alternatives and offer
some remarks in the concluding Section.

\section{The perturbative approach}

The {\sl ``perturbative" approach}, with ${\cal L}_0$ as the
free Lagrangian density, avoids the difficulties resulting from new poles in
the propagators, and takes into account the smallness of the breakings to
include them into the interaction Lagrangian density as super-renormalisable
couplings. The free photon and fermionic fields and their corresponding
asymptotic states are defined as usual. Moreover, the photon and electron
masses are defined by the same normalisation conditions as in ordinary QED,
$e.g.$
$$\dst <\psi(p)\bar{\psi}(-p)>^{prop.}\mid_{\not{p} = m\,,\;b = c =0}\quad =\
0\,,\quad\cdots$$ According to standard results in renormalisation theory, the
Lorentz invariance breaking adds new terms into the primitively divergent
proper Green functions. By power counting, these are
$$\Ga_{\mu\nu} (p,-p)\,,\
\Sigma (p,-p)\,,\ \Ga^{\rho} (p,q,-(p+q))\ {\rm and}\
\Ga_{\mu\nu\rho\si} (p_1,\,p_2,\,p_3,\,-(p_1+p_2+p_3))\,,$$
 respectively the photon and electron 2-points proper Green functions, the
photon-electron proper vertex function and the photon 4-point proper Green
function.  The corresponding overall  divergences (sub-divergences being
properly subtracted) are {\sl polynomial in the momenta and masses}\footnote{\
 C invariance has been used. The Ward identity (\ref{WI4}) will relate some of
these parameters : $a_2 = a_3 = a_4 = 0\,,\ a_{12} = e\ a_8\,,\ a_{13} = 0. $}:

\beqa\label{div1}\Ga_{\mu\nu}(p,-p)\mid_{div} & = & a_1[g_{\mu\nu} p^2 -
p_{\mu}p_{\nu}] + a_2 p_{\mu}p_{\nu} + [a_3 m^2 + a_4
\la^2]g_{\mu\nu} +\nnb\\ & + & [a_5 b^{\rho} + a_6
c^{\rho}]\epsilon_{\mu\nu\rho\si}p^{\si} + a_7 b^\mu b^\nu\,,\nnb\\
\Sigma(p,-p)\mid_{div} & = & a_8 \not{p} + a_{9} m + [a_{10} b^{\rho} + a_{11}
c^{\rho}]\ga_{\rho}\ga^5 \,,\\
\Ga^{\rho}(p,q,-(p+q))\mid_{div} & = & a_{12} \ga^{\rho} \quad : \quad {\rm no\
} b^\rho\ {\rm or\ } c^\rho\ {\rm dependance}
\,,\nnb\\
\Ga_{\mu\nu\rho\si}(p_i)\mid_{div} & = & a_{13} [g_{\mu\nu}g_{\rho\si} +
g_{\mu\rho}g_{\nu\si} + g_{\mu\si}g_{\rho\nu}] \quad : \quad {\rm no\
} b^\rho\ {\rm or\ } c^\rho\ {\rm dependance}\,.\nnb\eeqa
 All  parameters
$a_i$, positions
and residues of the poles in propagators, couplings at zero momenta,... - but
for the unphysical, non renormalised ones [such as the longitudinal photon
propagator (gauge parameter $\al$) and the photon regulator mass
$\la^2$ for unbroken QED] - require normalisation conditions, a point which
has often been missed since the successes of {\sl minimal} dimensional
regularisation scheme
\cite{GB1990} but is stressed in some reviews
\cite{{Victoria},{Chen2001}}.
 In particular we shall require 2 new normalisation conditions to fix the
breaking parameters
$b^{\mu}$ and
$c^{\nu}\,:$
\beqa\label{nor1} b^{\mu} &  = & - \frac{i}{4}
Tr[\ga^{\mu}\ga^5<\psi(p)\bar{\psi}(-p)>^{prop.}]\mid_{p=0}\,,\nnb \\
  c^{\mu} & = &
\frac{1}{12}\epsilon^{\mu\nu\rho\si}\frac{\partial}{\partial p^{\si}}
<A_{\nu}(p)A_{\rho}(-p)>^{prop.}\mid_{p=0}\,.\eeqa

\noindent Note that, contrary to  ${\cal L}_{int}(x)$ and ${\cal L}_1(x)\,,$
the
${\cal L}_2(x)$ term also breaks the local gauge invariance of the Lagrangian
density. But we emphasize the fact that - except for the unphysical part
$\dst\int [ -\frac{1}{2\al}(\partial A)^2 +
\frac{1}{2}\la^2 A_{\mu}^2 ] $ - the action
$\Gamma = \int \cal{L}$ is invariant under local gauge transformations.

So a Ward identity may be written  :
\beqa\label{WI4}
\lefteqn{\int d^4 x
\left\{\frac{1}{e}\partial_{\mu}\La(x)\frac{\de\Ga}{\de A_{\mu}(x)}+
i\La(x)[\bar{\psi}(x)\frac{\stackrel{\rightarrow}{\de}\Ga} {\de\bar{\psi}(x)} -
\frac{\Ga\stackrel{\leftarrow}{\de}}{\de{\psi}(x)}\psi(x)]\right\} =}
\nnb
\\  & =& \int d^4 x \left\{-\frac{1}{e\al}
\partial_{\mu}A^{\mu}(x)\Box\La(x)  +
\frac{\la^2}{e}A^{\mu}(x)\partial_{\mu}\La(x) +
\frac{1}{2e}\epsilon_{\al\be\de\mu}c^{\al}F^{\be\de}(x)
\partial^{\mu}\La(x)\right\}
\nnb
\\ & \Rightarrow & W_x\,\Ga\, \equiv \partial_{\mu}\frac{\de\Ga}{\de
A_{\mu}(x)} - ie[\bar{\psi}(x)\frac{\stackrel{\rightarrow}{\de}\Ga}
{\de\bar{\psi}(x)} -
\frac{\Ga\stackrel{\leftarrow}{\de}}{\de{\psi}(x)}\psi(x)] =
\frac{1}{\al} [\Box +
\al\la^2]\partial_{\mu}A^{\mu}(x)\,.
\eeqa 
We emphasize the fact that this equation is exactly the same as the one for
ordinary QED.

As soon as we use a regularisation that respects the symmetries (gauge,
Lorentz covariance and charge conjugation invariance), the perturbative proof
of renormalisability reduces to the check that the
${\cal{O}}(\hbar)$ quantum corrections to the classical action $\Ga\ :
\Ga_1 = \Ga|_{class.} + \hbar\De\,,$ constrained by the Ward identity
(\ref{WI4})  may be reabsorbed into the classical action through suitable
renormalisations of the fields and parameters of the theory. This has been
proven in \cite{Bonneau01}.

There, some local sources have been introduced to define the {\bf local}
operators ${\cal L}_1(x)$ and ${\cal L}_2(x)\,.$ Although this is only a
technical tool, it has been criticised
\footnote{\ For example in page 3 of
\cite{Chen2001} : {\it `` Bonneau introduced external source fields for the
axial vector current and the CS term, so the Ward identities he derived
actually impose gauge invariance on Lagrangian density ..."} This assertion is
wrong as the Lagrangian density is not gauge invariant (moreover it has been
gauge-fixed..) but, as proven in our analysis
\cite{Bonneau01}, the breaking of local gauge invariance {\bf is a soft one}
and may be seen as a complementary part in the gauge fixing, then
non-renormalised.} and we shall discuss later on this point.

Then in \cite{Bonneau01}, we have proven that, {\sl being linear in the quantum
field}, the variation of ${\cal L}_2(x)$ in a local gauge transformation is
soft : no essential difference occurs between local  gauge invariance of the
action and the ``softly" broken local gauge  invariance of the Lagrangian
density. As a consequence, the theory (\ref{QED1}) is consistent (even with no
${\cal L}_2(x)$ term) and the CS term has been shown to be unrenormalised,
{\bf to all orders of perturbation theory}. So, its experimental ``vanishing"
offers no constraint on the other CPT breaking term
${\cal L}_1(x)\,.$
\noindent To summarize, we have proven that : \begin{itemize}
\item The local gauge invariance of the Lagrangian density  is  destroyed by a
${\cal L}_2$ term (plus of course by the usual gauge fixing term) : but, being
bilinear in the gauge field, ${\cal L}_2(x)$ behaves as a minor modification
of the {\sl gauge fixing term} as
$\partial_{\nu}A^{\nu}$ remains a free field. As part of the ``gauge term",
this ${\cal L}_2(x)$ is,  as usual, not renormalised : so its all-order value
is equal to its (arbitrarily chosen) classical one.
\item In \cite{Costa} we also check that gauge invariance is respected
at second order in the breaking parameter $b^\mu\,.$
\item A theory with a vanishing tree level Chern-Simons like breaking term is
consistent as soon as it is correctly defined : thanks to the gauge invariance
of the action, we have proven that the normalisation condition
$c^{\mu} = 0$ may be enforced to all orders of perturbation theory.

\item The 2-photon Green function receives definite (as they are finite by
power counting) radiative corrections \cite{Bonneau01} $$\quad \simeq
\frac{\hbar e^2}{12\pi^2}
\frac{p^2}{m^2} \epsilon_{\mu\nu\rho\si} \,p^{\si}b^{\rho} + \cdots
$$ Recall the case of the electric charge : physically measurable quantities
occur only through the
$p^2$ dependence of the photon self-energy (as the Lamb-shift is a measurable
consequence of a non-measurable charge renormalisation). Unfortunately, as
Coleman and Glashow explained, the absence of birefringence of light in vacuum,
$i.e.$ the vanishing of the parameter $c^{\mu}\,,$ gives no constraint on the
value of the other one $b^{\mu}\,.$
\end{itemize}

\section{The non-perturbative approach}

The second solution, the ``{\sl
non-perturbative approach}'', introduces new poles in the fermion propagator
and requires a thorough discussion about causality and stability. Many papers
discuss that question
\cite{{Adam1},{Adam}}.

However, in the first paper in \cite{ColKos}, Colladay and Kosteleck{'}y gave
a direct analysis of the {\bf complete} classical fermion Green function  as
defined by ${\cal L}_0 + {\cal L}_1\,.$ In particular they check that the 
anticommutator of two fermionic fields vanishes for space-like separations, in
agreement with microcausality (at least for a time-like breaking $b_{\mu}\,).$
This confirms our analysis on the correctness of a theory with no classical CS
term. Then, Adam and Klinkhamer show that the addition of a ( radiatively
generated) CS term
${\cal L}_2(x)\,$  with a time-like $c^{\mu}$ breaks microcausality
\cite{Adam1}. As our non-renormalisation theorem ensures that, if absent at the
classical level, the CS term will not appear in higher-loop order,
 microcausality will not be destroyed in higher-loop order.

According to
\cite{Adam}, a vanishing Chern-Simons's like parameter $c^{\mu}$ is required
; for other analyses, only a time-like $b^{\mu}$ is allowed
 and a space-like $c^{\mu}$ \cite{{Adam},{Soldati1}}\footnote{ As the
radiatively generated $c^\mu$ is proportional to $b^\mu,$ if $c^\mu$ is
absent at the classical level, it is hard to understand how a time-like
$b^{\mu}$ 
 and a space-like $c^{\mu}$ can be coherent ?}. But none of
those papers were able to {\bf prove} the finiteness of the one-loop
corrections. On the contrary, note that if, as in the previous
``perturbative" case, one introduces local sources to define the local
operators 
${\cal L}_1(x)$ and
${\cal L}_2(x)\,,$ {\sl the one-loop finiteness will be obtained}, but as we
shall see on equ.(\ref{soldati}), the calculation algorithm does not respect
gauge invariance and finite ${\cal O}(\hbar)$ terms have to be added in the
Lagrangian density (see a discussion on another gauge invariance breaking
algorithm in
\cite{Costa}).

Anyway, the analyticity argument of Coleman and
Glashow is no longer at hand as the fermion propagator has new poles. Then,
it is not surprising that a non-vanishing induced Chern-Simons like term
appears.

To understand the discrepancies, let us discuss the electron propagator in both
situations :
\begin{itemize}
\item I) in the ``{\sl perturbative}" approach, the fermion two point function
is written as :
\beq\label{fermion1} i[\not p - m - \not b \ga^5]^{-1} = S^1_F(p) =
\sum_{n=0}^{\infty}\frac{i}{\not p - m}\left\{-i\not b \ga^5 \frac{i}{\not p -
m}\right\}^n\,;\eeq
\item II) in the ``{\sl non-perturbative}" approach, the fermion propagator is
taken as a whole :
\beqa\label{fermion2} i[\not p - m - \not b \ga^5]^{-1} = S^{2-a}_F(p) &
\stackrel{\cite{Soldati1}}\equiv & i\frac{p^2 -m^2 + b^2 +2(b. p +m\not
b)\ga^5}{(p^2 -m^2 + b^2)^2 -4[(b\dot p)^2-m^2 b^2]}(\not p +m +\not b
\ga^5)\,,\\ = S^{2-b}_F(p) & \stackrel{\cite{Victoria99}}\equiv & i(\not p +m
-\not b
\ga^5)\frac{p^2 -m^2 - b^2 +[\not p,\not b]\ga^5}{(p^2 -m^2 + b^2)^2 -4[(b.
p)^2-m^2 b^2]}  \,.\nnb\eeqa
These two equations (\ref{fermion2}) illustrate two of the possible
equivalent expressions for the complete fermionic propagator.
\end{itemize} 

Of course, $b$ being a very small parameter, $S^2_F(p)$ may be
expanded in power-series, and one recovers $S^1_F(p)\,.$ 

\noindent However, expanding in powers of $b$ from the very beginning, or at
the end of the calculation of the Green function, makes some
difference due to the question of regularisation (recall that when a Green
function is primitively divergent, the Feynman integral should be regularised
before any manipulation), and of the choice of the computational algorithm for
the Green functions.

Let us consider dimensional regularisation with the unique
consistent formulation in presence of the $\ga^5$ matrix, the one of t'Hooft
systematized by Breitenlohner and Maison \cite{{t'Hooft},{DR1}} (for a review
see
\cite{GB1990}) :
\beqa\label{dimreg}
\ga^{\mu} = \hat{\ga^{\mu}} + \hat{\hat{\ga^{\mu}}}\,,\quad
\{\ga^{\mu}, \ga^{\nu}\} = 2g^{\mu\nu} &,\quad g^{\mu}_{\mu} = D &,\quad
\hat{g}^{\mu}_{\mu} = D-4\,,\quad \hat{\hat{g}}^{\mu}_{\mu} = 4\,,\\
\{\hat{\hat{\ga}}^{\mu}, \hat{\ga}^{\nu}\} =0\,,
\{\ga^5, \hat{\hat{\ga^{\mu}}}\} = 0 & ,\quad [\ga^5, \hat{\ga^{\mu}}] = 0
&, Trace[\ga^5
\hat{\hat{\ga}}^{\mu}\hat{\hat{\ga}}^{\nu}\hat{\hat{\ga}}^{\rho}\hat{\hat{\ga}}^{\si}]
= 4\epsilon^{\mu\nu\rho\si}\,,\quad e.t.c.... \nnb
\eeqa

\noindent One should not be surprised that extensions to $D$ dimensions of the
different expressions [equs.(\ref{fermion1}), (\ref{fermion2}a - b)] will lead
to different results, differences being given by``evanescent operators" (which
vanish when $D$ goes to 4 ) but giving finite contributions when inserted into
{\sl a priori} primitively divergent graphs. This illustrates the second
reason for discrepancies among published results. To be more definite, one can
easily show that \footnote{\ Notice that, in a one-loop calculation, the
physical $b$ parameter and the external photon momentum and component indices
$\mu,\nu,..$ stay in $D=4$ dimensions : $b^{\mu} \equiv \hat{\hat
{b}}^{\mu}\,, \{\ga^5,
\not b\} = 0\,.$ } :
\beq\label{evanesc1}  S^1_F(p)  = S^{2-a}_F(p) + 2\hat{\not p}\not
b\frac{\ga^5(\not{p} +m)}{(p^2-m^2)^2} + 2b^2\frac{(\not p +m)\hat{\not
p}(\not p +m)}{(p^2-m^2)^3} + {\cal O}(b^3)\,,\eeq

and

\beq\label{evanesc2}  S^1_F(p)  = S^{2-b}_F(p) + 2\hat{\not
p}\ga^5\frac{(\not{p} +m)\not b}{(p^2-m^2)^2} - 2b^2\frac{(\not p +m
-4\hat{\not p})\hat{\not p}(\not p +m)}{(p^2-m^2)^3} + {\cal O}(b^3)\,.\eeq

Notice that these evanescent terms do not come from taking the inverse `` in
$D$ dimensions'' of the proper ``non-perturbative" Green function $i[\not p - m
-
\not b
\ga^5]$ as claimed in
\cite{Soldati1}, but rather from {\bf different continuations in
$D-$dimensions} of various $D=4$ identical quantities obtained either
from``non-perturbative" or ``perturbative" approaches.

The complete one-loop calculation (up to third order in $b^{\mu}$) of the
photon vacuum polarisation tensor with the ``perturbative" ($i.e.$ with
$S^1_F(p)$ of equ.(\ref{fermion1}) as fermion propagator), gauge invariant
approach and consistent dimensional regularisation or Pauli-Vilars method, may
be found respectively in
\cite{{Bonneau01},{Costa}} where it is shown that
\beqa\label{bonneauCosta}
\Pi^{\mu\nu}(p,-p) & = & i \frac{ e^2}{12 \pi^2}[g^{\mu \nu}p^2 -  p^{\mu}
p^{\nu}] \left\{ \log\frac{4\pi\mu^2}{m^2}  -  p^2 \int_0^1 dz \;\frac{[1- 2 Z
- 8 Z^2]}{2\Delta} \right\} + \nnb \\ & + & i\frac{e^2}{2 \pi^2}\epsilon^{\mu
\nu
\alpha
\beta} p_{\alpha} b_{\beta } \left\{ p^2 \int_0^1 dz\;
\frac{Z}{\Delta} \right\}  +i \frac{e^2}{\pi ^2} X^{\mu\nu} 
\int_0^1 dz \left[ \frac{Z}{\Delta}  + \frac{ Z^2}{[\Delta]^2} p^2
\right] \,, \eeqa
 where : \begin{itemize}\item$\mu$ is the UV scale needed to renormalize the
electric charge,\item
$Z=z(1-z)$ and $\Delta = m^2 -Z p^2\,,$ \item  $ X^{\mu\nu}$ is the unique
polynomial tensor of canonical dimension 4, quadratic in
$b^{\alpha}$ and transverse with respect to $p^\mu$ and $b^\mu$, $$ X^{\mu\nu} 
=  b^2 (g^{\mu\nu}p^2 - p^{\mu}p^{\nu})  - g^{\mu\nu}(p.b)^2 - p^2
b^{\mu}b^{\nu} + (p.b)(p^{\mu}b^{\nu} - p^{\nu}b^{\mu}) \,.$$
\end{itemize}

\noindent On the other hand, in \cite{Soldati2} , the calculation (still up to third
order in
$b^{\mu}$ and in the limit $p^2 \rightarrow 0$) with the fermion propagator
$S^{2-a}_F(p)$ of equ.(\ref{fermion2}) gives
\beq\label{soldati}
\Pi^{\mu\nu}(p,-p)  =  i [g^{\mu \nu}p^2 -  p^{\mu} p^{\nu}] \Pi_{div}(0) +
i\frac{e^2}{2
\pi^2}\epsilon^{\mu
\nu
\alpha
\beta} p_{\alpha} b_{\beta } +i \frac{e^2}{6\pi ^2}\left[g^{\mu \nu}b^2
+\frac{1}{m^2} X^{\mu\nu} \right]
 \,. \eeq

As $p_\mu \Pi^{\mu\nu}(p,-p) \neq 0\,,$ gauge invariance is lost. However,
if one computes the extra contributions coming from the ``evanescent" terms
in equ.(\ref{evanesc1}), one obtains  (still in the limit
$p^2
\rightarrow 0$)~:
\beq
\Delta \Pi^{\mu\nu}(p,-p) =  -i\frac{e^2}{2
\pi^2}\epsilon^{\mu\nu\alpha\beta} p_{\alpha} b_{\beta } -i \frac{e^2}{6\pi
^2}g^{\mu \nu}b^2  \,,
\eeq which corresponds exactly to the difference between
equ.(\ref{bonneauCosta}) and equ.(\ref{soldati}) (see also the remark in
appendix B of \cite{Soldati1}).

To summarize, the so-claimed ``unambiguous" results in
\cite{{Soldati1},{Soldati2}} do not come from enforcing some ``maximal
residual symmetry" at the quantum level since we obtained the same result as
theirs by using dimensional regularisation and the modified propagator
(\ref{evanesc1}). Moreover, if the approach of a physical cut-off in the three
dimensional momentum space for fermions ( as developped in
\cite{Soldati1}) is physically interesting, actually it plays no role in their
computation. 

\noindent Indeed, as we now explain, the sole ingredient of their calculation
is a specific choice of regularisation. 
\begin{itemize}\item First, consider the computation
in [\cite{Soldati1}, equ.(4.10)] of
\beq\label{equa1}\dst\int
\frac{d^4 p}{(2\pi)^4}\frac{b_{\rho}(p^2+3m^2) - 4p_{\rho}(b.p)}{(p^2-m^2+i
\epsilon)^3} \ :\eeq 
\begin{itemize}\item in \cite{Soldati1}, a purely time-like
$b\equiv (b_0,0,0,0)
$ is chosen and the authors firstly integrate on the variable  $p_0$ (no need
of a cutoff as the integration happens to be possible between $-\infty$ and
$+\infty$). After that, the integration on the three dimensional momentum
space variables also converges and gives the announced result
$i/(2\pi^2)\,,$ without any need to refer to an ``SO(3)" residual
symmetry ;
\item however, if one firstly integrates on the three dimensional momentum
space variables, which also happens to be convergent, one finds a vanishing
result, before the (convergent) $p_0$ integration ;
\item then, this proves
(recall Fubini's theorem) that the ``four dimensions'' integral (\ref{equa1})
does not exist as a multidimensional one, even if it happens to be finite
(which is still the main point to be understood !) ;
\item finally, one easily checks that $D$ dimensional continuation of
(\ref{equa1}) and integration with average of
$p_\rho p_\la
\equiv p^2g_{\rho\la}/D$ gives the same result $i/(2\pi^2)\,.$
\end{itemize}

\item In the same manner, in Section 5 of the same paper, the authors in fact
use  dimensional regularisation and ``$D$ dimensional spherical coordinates'' :
here again, their calculation does not rely upon the
claimed ``maximal residual symmetry". 
\end{itemize}

\noindent {\bf Remark :} of course, as the authors remarked, if averaging is
done in 4 dimensions - which destroys the gauge invariance of the dimensional
regularisation scheme \footnote{\,In particular, this will modify the ordinary
QED vacuum polarisation tensor and give a quadratic divergence !} (see
\cite{t'Hooft}-p.196,\cite{GB1990}) -, a different result is obtained
\cite{{Jackiw},{Victoria}}. This illustrates the second reason for
discrepancies between published results.

\section{Discussion and concluding remarks }

As a complement to the reviews
\cite{{Victoria},{Chen2001}}, let us now comment upon some points given in the
literature :

\begin{itemize}\item {\bf Use of local sources in the classical action} 

As said before, the use of local sources in \cite{Bonneau01} has been
criticized. Of course, in ordinary QED, the axial current, being uncoupled, is
absent from the Lagrangian density and so does not need to be defined as a
quantum operator ; no axial vertex being present, {\it a fortiori} there is no
axial anomaly and no triangle graph to consider. However, as soon as PCAC is
used to compute the decay
$\Pi^0 \rightarrow \ga\ \ga \,,$ this triangle graph has to be computed and
gauge invariance of the unintegrated three-vertices function is required.

On the contrary, in CPT-broken QED of equ.(\ref{QED1}), new axial insertions
enter the game, but they are integrated ones $\dst\int d^4x{\cal L}_1(x)$ and
$\dst\int d^4x{\cal L}_2(x)\,.$ Then, some authors argue that introducing local
sources for the Lagrangian density breakings means adding supplementary
conditions on the theory. However, as explained in footnote 3, no local gauge
invariance hypothesis results from the introduction of local sources for these
insertions.

Suppose that one has only the
Ward identity (\ref{WI4}) at hand to constrain the possible
ultra-violet divergences of equ.(\ref{div1}). This is not sufficient to
prove that the breakings introduce no new infinities : in
particular, the Chern Simons term is of the right canonical
dimension and quantum numbers and
 satisfies the Ward-identity (\ref{WI4}) :
$$p^{\mu}\Ga_{\mu\nu}(p,-p) = 0 \quad \stackrel{ in\
particular}{\Rightarrow}  p^{\mu}<[\dst\int d^4x{\cal L}_1(x)]
A^{\mu}(p) A^{\nu}(-p)> = 0\,.$$ So, first, we have no explanation of
the fact that all one-loop calculations of the CPT breaking contribution
to the photon self-energy give a  finite result ($a_5 = a_6 = a_7 = 0$),
 second, {\sl being unconstrained, its finite part} (renormalised value)
has to be fixed by a normalisation condition (a different situation
than a radiative correction such as the (g-2) or the Lamb-shift for
example). So, no prediction is possible and its value remains
arbitrary, which is rather unsatisfactory. 

\item {\bf Use of the heat-kernel expansion}

In \cite{Sitenko01},
the one-loop calculation of the CS correction is done with the heat-kernel
expansion and the Schwinger proper-time method, leading to a new finite
result, claimed to be unambiguously determined. However,
\begin{itemize}\item  here again there is no explanation of the absence of
infinities in the result : then the finite part is {\sl a priori} ambiguous
\footnote{\ Remember that in Fujikawa 's calculation of the axial anomaly, 
gauge invariance was implemented through the basis used to compute the
fermionic Jacobian : he chose eigenvectors of the operator
$i\not{\partial} -e\not{A}\,;$  another choice would allow the transfer of the
axial anomaly to some vector anomaly (see also the discussion on the ``minimal
anomaly" in non-abelian gauge theory) \cite{Fujikawa}.},
\item other computations with the Schwinger proper-time method exist
\cite{proper-time} and give a different result, proving at least that some
``ambiguity" remains,
\item some terms are lacking in this calculation : in particular a
logarithmically divergent contribution to the CS term results from a thorough
computation of the quantity given in equation (21) of
\cite{Sitenko01} ( in the absence of any precise criteria to substract
infinite parts, this should not be a surprise).
\end{itemize}

\end{itemize}
\vspace{1cm}

In that work, we traced the main origin of the controversy on a possible
Chern-Simons like term generated in PCT- broken QED (and on the $b^2$
contribution to the vacuum polarisation tensor) back to the delicate choice of
the ``unperturbed" Lagrangian density.

For us and other authors \cite{{Adam1},{Adam}}, the {\sl non-perturbative}
choice suffers from delicate theoretical problems (microcausality, analyticity
...) ; moreover, in the absence of any Ward-identity \footnote{\ In
particular, we proved that the claimed ``maximal residual symmetry" is in fact
not used in actual computations, which then rely in some blind way upon
regularisation (even on the order  of the integration in multidimensional
integrals ..).} (we emphasized that, without introduction of local sources for
the breaking terms, the gauge invariance of the complete action cannot be fully
exploited), one is unable to {\sl explain} the main phenomenum~: the
finiteness of all results, which allows for an unambiguous prediction.
Moreover, as in that context the finiteness would appear as an ``accidental
one", there would be no reason that such result holds to higher loop order.

\bibliographystyle{plain}

\end{document}